\documentclass[a4paper]{mem}
\usepackage{natbib}
\usepackage{graphicx}
\usepackage[a4paper]{hyperref}
\idline{73}{23}
\begin{document}
   \title{X-ray properties of the ``Composite'' Seyfert/Star-forming galaxies
}

   \author{A. Wolter\inst{1}, G. Trinchieri\inst{1}, R. Della Ceca\inst{1}, 
F. Panessa\inst{2,3}, M. Dadina\inst{2}, L. Bassani\inst{2}, M. Cappi\inst{2}, A. Fruscione \inst{3}, S. Pellegrini\inst{4}, 
         \and
G. Palumbo \inst{4}  
}

   \offprints{A. Wolter, OAB}
\mail{via Brera, 28, 20121 MILANO }

   \institute{Osservatorio Astronomico di Brera
via Brera 28, 20121 MILANO  \email{anna@brera.mi.astro.it}\\ 
              \and  INAF/IASF, via P. Gobetti 101, 40129 Bologna  \\
              \and Center for Astrophysics, 60 Garden St., Cambridge, MASS\\
              \and Dipartimento di Astronomia, Universit\`a di Bologna, 
via Ranzani 1, 40127 Bologna
             }

   \abstract{
An enigmatic, small class of IR and X-ray luminous sources, named
``Composite'' starburst/Seyfert galaxies, has been defined from IRAS
and RASS data.  The objects have optical spectra dominated by the
features of HII galaxies (plus, in some cases, weak Seyfert
signatures) but X-ray luminosities higher than expected from starbursts
and more typical of Seyfert nuclei.  The true
nature of this class of objects is still unknown.  We present
Chandra data of four of these galaxies that were obtained to investigate the
nature of the X-ray source.
The X-ray spectrum, the lack of any significant extended component,
and the observed variability indicate that the AGN is the dominant
component in the X-ray domain.
  
} 

\authorrunning{A. Wolter et al.}  
\titlerunning{``Composite'' Seyfert/Star-forming galaxies } 
\maketitle
%

\section{Introduction}

An enigmatic class of six low redshift galaxies has been defined
during a spectroscopic survey of bright IRAS and RASS sources \citep{moran96}.
They were named ``Composite'' to indicate their dual
nature: the optical spectrum typical of a star forming galaxy is
characterized by additional weak broad H$\alpha$ wings that, together with
the high X-ray luminosities, suggests the presence of an AGN.
We have performed a study with Chandra of the four ``Composites'' that
still lack a high resolution X-ray observation, with two main goals:
1) to investigate any extended X-ray component possibly associated with
    the InfraRed Starburst; 
2) to determine the nature of the nuclear
    emission.
We present here the first results obtained, while a detailed description
of the X-ray data will appear in Panessa et al. (in preparation).

\section{X-ray data}

The sources were all observed with the ACIS-I instrument on board Chandra
for a nominal exposure of $\sim 25$ ksec. Relevant data for the sources are listed 
in Table~\ref{tab}.
Based on the RASS fluxes we had requested a Chandra
configuration that would allow more frequent read-out of the data 
by using a smaller field of view to avoid pile-up.
The results however indicate that the precautions taken were not necessary,
since the observed count rates (see Table~\ref{tab}) are well below the
expected value and the pile-up is not significant.

\begin{table*}
\caption{Properties of the 4 ``Composite'' galaxies under study.}
\begin{tabular} {| l r c r c c|}
\hline
Name      &    Redshift &  c/s & R$^*$ &  L$_X$(2-10keV)  &  L(FIR) \\
        &          & ACIS-I &   & cgs (Chandra) & cgs \\
\hline
IRAS01072+4954  & 0.0237 &0.063$\pm$0.002 &0.5 &$3.85 \times 10^{41}$& $1.22 \times 10^{44}$\\
IRAS01319--1604 & 0.0199 &0.115$\pm$0.002 &0.5 &$6.99 \times 10^{41}$& $1.97 \times 10^{44}$\\
IRAS04392--0123 & 0.0289 &0.015$\pm$0.001 &0.1 &$3.68 \times 10^{41}$& $1.81 \times 10^{44}$\\
IRAS20069+5929  & 0.0374 &0.180$\pm$0.003 &1.0 &$6.84 \times 10^{42}$& $7.52 \times 10^{44}$\\
\hline
\end{tabular}

$^*$ R is the ratio between Chandra and RASS fluxes. \\
\label{tab}
\end{table*}

\smallskip
\subsection{Variability}

   \begin{figure}
   \centering
   \includegraphics[width=6cm]{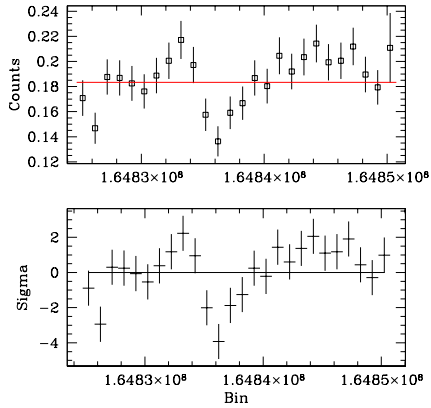}
   \caption{The light curve of IRAS20069+5929, in bins of 1000 sec, and
residuals with respect to a constant (in $\sigma$); the fit $\chi^2$ is 
46.8/25, indicating that variability is present. }
              \label{lc}%
    \end{figure}

We compare the total Chandra luminosities to previous RASS data 
\citep{moran96}. We find consistency only for IRAS20069+5929.
The luminosity is smaller by a factor $\sim$ 10 for IRAS04392-0123, and
a factor $\sim$ 2 for IRAS01072+4954 \& IRAS01319-1604, as indicated
by R in Table~\ref{tab}. 
A discrepancy of a factor $\sim$ 25 was also reported for IRAS00317-2142 
\citep{georga03}, while no information is available so far for IRAS20051-1117
that has been observed by Chandra and XMM-Newton.
The long term variability alone already indicates that the dominant
source of emission is due to a point sources. 

We have further investigated short term variability.
We compute light curves by extracting the source counts in a region of
4$^{\prime\prime}$ radius and background in a larger, source free, annulus 
around it.
We bin light curves at 1000 sec to have enough statistics per
bin.  We show in Figure~\ref{lc} the non constant light curve
of source IRAS20069+5929, which has the largest statistics of the 4 Chandra 
observations. 
We highlight this variability detection since it indicates the presence of 
a point source even if the average luminosity of this 
source has not changed since RASS.

\smallskip
\subsection{Spatial distribution}

   \begin{figure}
   \centering
   \includegraphics[width=6cm,height=5cm,clip=]{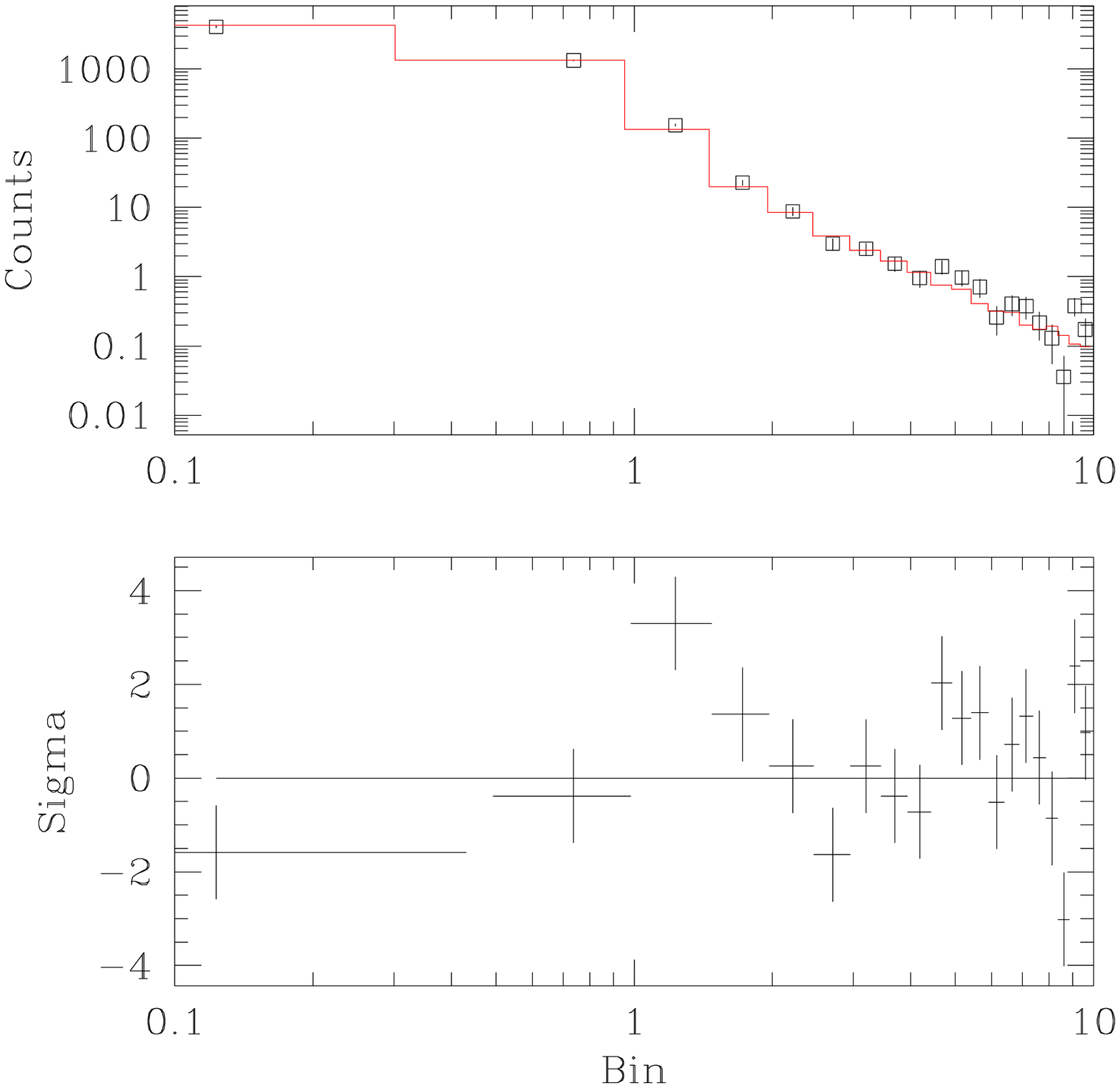}
   \caption{
Profile of IRAS20069+5929 (squares) compared to the 
{\it chart}-derived PSF with the same spectral distribution of the source 
(histogram). X-axis is in arcsec. 
}
              \label{psf}%
    \end{figure}

No extent is evident in any of the four sources, although a low surface
brightness component might be present; more detailed analysis in underway.
With the {\em ciao} task {\em chart} we simulate a point source distribution
of photons that has the same spectrum of the observed source. We then
extract the profile of both the source the Chandra PSF to compare them.
We plot the histograms in Figure~\ref{psf} for the brightest source 
IRAS20069+5929 and verify that it is consistent with a point source.

\smallskip
\subsection{X-ray spectra}

We extract spectra for the four sources and fit them with a simple model:
a power law modified by low energy
absorption due to our galaxy (from \citealt{dickey90}) plus an
intrinsic redshifted component such as that expected in the hypothesis that
the bright broad
component in the optical spectrum is not seen because of absorption.  We
find instead that intrinsic N$_{\rm H}$ is always small or negligible:
IRAS01072+4954 and IRAS01319-1604 do not require statistically an
intrinsic N$_{\rm H}$ component; 
IRAS04392-0123 has N$_{\rm H}^{intrinsic} \leq 1.17 \times 10^{21} cm^{-2}$;
IRAS20069+5929 has N$_{\rm H}^{intrinsic} = 2.69 [2.21-3.22]\times 10^{21} cm^{-2}$. 

A similar behavior is reported for IRAS 00317-2142, with ASCA data 
\citep{georga00}, where there is no evidence of intrinsic absorption above
a few $10^{20} cm^{-2}$.

   \begin{figure}
   \centering
   \includegraphics[width=6cm]{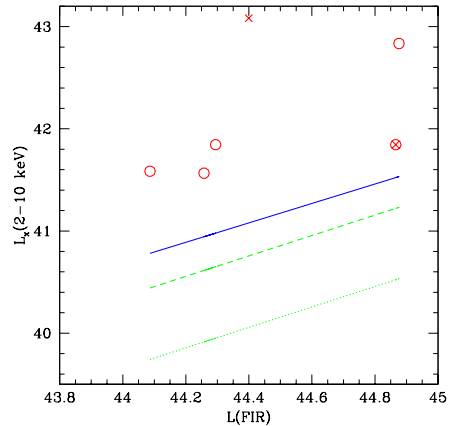}
   \caption{The X-ray hard luminosity from Chandra vs. the FIR
luminosity from \citet{moran96} (circles). 
The cross represents IRAS 20005-1117 (from \citealt{moran96}, X-ray from RASS);
circle+cross is IRAS 00317-2142 (from \citealt{georga03}, X-ray from Chandra).
Dotted and dashed lines from \citealt{persic04}, solid line from 
\citealt{david92}. See text for details. }
              \label{lxfir}%
    \end{figure}

In spite of the lower luminosities in the most recent
measurements, the current values are still not consistent
with the hypothesis of them being produced by the starburst seen at InfraRed
wavelengths. This is evident in Figure~\ref{lxfir}, where
we plot (circles) the X-ray hard luminosity from Chandra vs. the FIR
luminosity from \citet{moran96}. The cross represents IRAS 20005-1117 
(from \citealt{moran96}, X-rays from RASS); circle+cross is IRAS 00317-2142
(from \citealt{georga03}, X-rays from Chandra).  From the FIR
Luminosity we compute the expected Star Formation Rate 
(SFR = 8-80 M$_{\odot}$/yr).  From the SFR
we derive the expected X-ray luminosity according to \citet{persic04} 
(dotted line is contribution from HMXB; 
dashed line is the total from the Starburst) or \citet{david92} (solid 
line).  The currently observed X-ray luminosities
of the sources are still at least about a factor of 10 higher than both
estimates. The AGN dominates in X-rays or the starburst component is
anomalous.  

\section{Conclusions}

From the X-ray Chandra observations we conclude that 
the sources are:
1) point-like in the Chandra images, although we cannot exclude a low surface 
brightness component; 
2) variable - both within the single Chandra observation (as in the case
of IRAS20069+5929) and with respect to older data; 
3) a thermal component is in general not present, even if in a few
cases it might account for part of the soft emission;
4) the intrinsic absorption measured in the X-ray band is small or not
required by the fit;
5) the current L$_X$ is still at least an order of magnitude higher
than the predictions from the starburst power, which is of the order of 
SFR = 10-100 $M_{\odot}$/yr.

Since the emission lines from the AGN are weak in the optical band, and  
absorption is not detected in X-rays,  we cannot suggest obscuration as 
responsible for hiding the optical AGN.

Therefore we conclude that an AGN is present and dominant 
in X rays, but probably of low luminosity:  L$_X$(2-10 keV)=3.5$\times 10^{41}$-7.$\times 10^{42}$  erg/s. 
This could explain the optical faintness observed by \citet{moran96}.

\begin{acknowledgements}
We acknowledge partial financial support from Agenzia Spaziale Italiana (ASI).
\end{acknowledgements}

\bibliographystyle{aa}

\end{document}